# Theory of Quantum Gravity of photon confirms experimental results of a varying fine structure constant while Quantum Mechanics leads to String theory

Pradip Kumar Chatterjee

Indian Physical Society
IACS Campus,2A & 2B Raja Subodh Chandra Mullick Road,Calcutta 700032
India

## Abstract

Quantum Mechanics of photons leads to a theory of Quantum Gravity that nicely matches the experimental results of varying fine structure constant, obtained from many-multiplet Quaser absorption systems and atomic clocks. The variation of that constant is due to quantum gravity of photons, created by their non-zero invariant mass. The photon mass is obtained from a Klein-Gordon scalar tachyon. This led to a Lorentz symmetry-breaking and varying speed of light theory in complex spacetime manifold. In essence, Quantum Mechanics includes quantum gravitational potential in the guise of Quantum potential. The greatest surprise lies in showing that Quantum Mechanics naturally leads to open bosonic string whose troublesome tachyonic vibration is taken in its stride. Quantum Mechanics also proves Sen`s second conjecture and space-tearing. Length melts into dimensionless number at the Planck scale. Quantum-mechanical analog of the classical equation $E=Mc^2$ has been derived and a dispersion relation demonstrates Lorentz non-invariance in Quantum Mechanics.

# 1. INTRODUCTION

Theory of Quantum Gravity is perhaps the greatest intellectual challenge to physicists today. Although the three interactions in nature --- electromagnetic, weak and strong --- have been successfully described by Quantum Field Theories (that constitute the Standard Model), gravity has persistently resisted unification with Quantum theory. There are several distinct paths along which attempts have been made in the past [1], but the following two theories have emerged with a sustained growth and internal consistency; and their recent results have far-reaching consequences in the understanding of our physical world:

(1) Loop Quantum Gravity [2,3,4,5,6,7]
(2) String, Superstring, M-theory [8,9,10,11]

It is a real conundrum why gravity cannot be united with Quantum Mechanics. When Quantum Gravity is teased out of Quantum Mechanics in a deceptively simple way, its mathematics spontaneously leads to the theory of open strings [12] and also tachyon condensation [13], when photon dynamics is considered. It is really happenstance that quantum gravity has found its experimental verification in the recent diverse experiments aimed at finding variations of fine structure constant over different look-back times [14,15,16].The first hint of a varying fine structure constant surfaced in string theory. Since then there has been a surge of interest in finding observational and theoretical results in support of a varying $\alpha$. In 1999, J.K.Webb et al [14] declared results showing a time-varying $\alpha$. Peres [17] suggested that the variation in $\alpha$ is due to a varying speed of light (VSL). The controversy regarding the effect of a varying dimensionful quantity on a dimensionless constant has been discussed below, and a solution to this end emerged when one could identify quantum spacetime with complex spacetime manifold. The main purpose of this paper is to show that our theoretical predictions from the quantum gravity of photons agree quite nicely with the accumulated experimental results of a time-varying $\alpha$. The other aim is to access string theory through Quantum Mechanics.

Various theories have been advanced to interpret a varying $\alpha$ over cosmological times [18,19,20,21,22]. A varying $\alpha$ creates many troubles in present-day physics ---- chief among these being a contradiction of the Standard Model of particle physics. It however more than compensates for this by yielding a theory of quantum gravity of photons, nicely verified by the extant experimental results provided by many-multiplet quaser absorption systems, Gamma Ray Bursts(GRB) and transitions between two nearly degenerate states of atomic clock [14,15,16,23,24,25,26,27,28,29,30,31,32,33,34].

Another striking result is obtained from this quantum gravity theory in a form of mathematical gift : Quantum Mechanics,if probed deeply, leads to open string theory [12]. A three-dimensional vibration also requires the use of quaternions in Quantum theory [35,36].

The organization of the paper is as follows : In section 2, I derive the photon wave function by quantizing the classical energy equation of special relativity, and then refute all objections to a photon wave function. The momentum operator is found invertible and the eigenvalue of this inverse operator is just the reciprocal of eigenvalue of the operator. In section 3, I derive another wave function, but this time it is motivated by a purely mathematical reasoning : The derivatives $\frac{d\phi_1}{dx}$ and $\frac{d\phi_2}{dt}$ need not necessarily be constants; they might be functions of x and t respectively. This yields a space- and time-varying speed of light in complex spacetime manifold. The controversy regarding the validity of VSL theories, in the context of a varying dimensionless number $\alpha$, loses strength once it is realized that real-valued meter rods and real time-ticking clocks are of no use in measuring a complex-valued speed of photon traveling in complex spacetime. The varying speed of light extracted in real spacetime is found greater than the speed of light in vacuum, $c_0$. This immediately made us consider tachyons --- the dreaded objects with imaginary mass, and not detected to date. But I have made a significant change in their property : Their imaginary energy and momentum are retained. These are not new concepts in Quantum Mechanics embedded in complex spacetime (see Ref [37] for details). Tachyon mass is kept real. A tachyon cannot be found in complex spacetime because the latter is operationally inaccessible. It cannot be found at a measurement event [37] since it pops up as a photon there. While discussing these I have deduced an explicit expression for the relativistic mass operator $\hat{m}_r(v) = m_r(\hat{v})$.

Next I turned to the problem of finding the least possible invariant mass of a tachyon in section 4. To this end, an ansatz is inserted into Klein-Gordon equation to obtain a four-component wave function. Arguments have been advanced to take away the sting of all the objections to its use as a valid wave function. The non-zero mass of photon (or, its complex spacetime masquerader, tachyon) has been found to be $1.7 \times 10^{-38}$ gm which is exceptionally close to values obtained in GRB events. The quantum-mechanical analog of the classical relation $E = Mc^2$ now reads

$$E = Mc^2 - \frac{\hbar^2}{8M}\left[1 - \frac{v^2}{c^2}\right].$$

From the varying speed of light I derive the the most important relation concerning variation in fine structure constant, in section 5 :

$$\frac{\Delta\alpha}{\alpha} = -\lambda c_0 t = -\left(\frac{2\sqrt{2}}{\pi\sqrt{3}}\right)l_p c_0 t$$

where the look-back time is $t$ and the constant $\lambda$ was later found to be proportional to Planck length $l_p$. The above equation explicitly shows that a time-varying $\alpha$ occurs at Planck scale.

Section 6 describes the Quantum Gravity theory of photons derived from quantum potential that introduces slowing down of older photons. The key ingredient in quantum gravity of photons is their non-zero invariant mass revealed by the theory. An important finding is that the Schwarzschild radius for photon is equal to $l_p^2$. There is indication in the theory that $\alpha'$ of string theory is also equal to $l_p^2$. As a refinement of Heisenberg`s position-momentum uncertainty relation we prove a theorem stating that a particle cannot have a definite position; nor it can have a definite momentum. Complete uncertainty ! The theorem in general states that an incompatible observable cannot have a definite value.This introduces a formidable tool in quantum physics : It immediately removes big bang singularity of Quantum Cosmology, singularities owing to notorious zero-distance interactions in QED, and Schwarzschild black-hole singularity.

I derive tachyon wave function in this section. The half wave number along the imaginary axis drove us to the idea that a single vertical wave is confined in an inaccessible region of Planck length. This oscillation is traveling with a velocity c(x,t) along the real x-axis. This describes an open string whose lowest excitation produces tachyons. I have shown that tachyons in complex space-time show up as photons at the measurement point. This is equivalent to tachyon condensation [38,39] which carries the tachyon to a stable state, viz., photon at the measurement point (MP).

In the same section we discuss the possible Lorentz non-invariance owing to non-zero invariant mass of photon. While the (classical) theory of special relativity does not admit a preferred reference frame, the measurement problem of Quantum Mechanics[40] poses the problem of preferred basis[41] which can be resolved only by stating that the preferred reference frame(the frame containing the device particle at the MP as the origin) is chosen by the observer or measuring device[37].This is also supported by Copenhagen Interpretation[40]. Quantum Mechanics requires ten-dimensional spacetime manifold [37] to describe a quantum system. Of these, five are imaginary dimensions which are obviously compactified since measurements always take place at real spacetime.The negative norm states of string theory are no longer disastrous as Quantum Mechanics permits negative probability[37]. String theory is therefore on the right track !

Finally, section 6 compares the predictions of our quantum gravity theory of photons with the experimental results gleaned from observational data of recent years showing the persistent variations $\dfrac{\Delta\alpha}{\alpha}$ and $\dfrac{1}{\alpha}\dfrac{d\alpha}{dt}$. These include the quaser and atomic clock experiments. The theory is in good agreement with the experimental results. This perhaps fulfils a hope that a quantum gravity theory has been experimentally confirmed. The experimental results have been provided with a consistent theoretical underpinning that stems from Quantum Mechanics.

# 2. PHOTON  WAVE  FUNCTION

There are clear-cut objections to a photon wave function [42A,43,44]. I shall discuss them only after deriving the wave function. The energy equation of special theory of relativity is

$$E^2 = p^2 c^2 + m^2 c^4$$

For a photon, the invariant mass $m$ is zero. Hence the quantized form of the above equation is

$$\hat{E}^2 \psi = \hat{p}^2 c^2 \psi \qquad (1)$$

Being an operator equation, the above equation cannot be simplified to

$$\hat{E}\psi = \hat{p}c\psi.$$

Replacing $\hat{E}$ and $\hat{p}$ by their explicit forms, Eq.(1) reads

$$\frac{\partial^2 \psi}{\partial t^2} = c^2 \frac{\partial^2 \psi}{\partial x^2}. \qquad (2)$$

The concept of wave function of photon is not new, but it has only found reference within the description of second quantization with creation and annihilation operators [42]. For the wave function of photon, we assume the splitting:

$$\psi(x,t) = \psi_1(x)\psi_2(t)$$

Since $\psi(x,t)$ is complex, there is no reason to believe that any one of $\psi_1(x)$ or $\psi_2(t)$ will be real. In general, both will be complex. Therefore,

$$\psi_1(x) = R_1(x)\exp[i\phi_1(x)],$$

where $P_1(x) = |\psi_1(x)|^2 = R_1^2(x)$. Also,

$$\psi_2(t) = R_2(t)\exp[i\phi_2(t)]$$

where $P_2(t) = |\psi_2(t)|^2 = R_2^2(t)$. Consequently, the ansatz for photon wave function is

$$\psi(x,t) = R_1(x)R_2(t)\exp[i\phi_1(x) + i\phi_2(t)]. \qquad (3)$$

The probability density is

$$P(x,t) = P_1(x)P_2(t) = R_1^2 R_2^2. \qquad (4)$$

The definitions of wave number $\kappa$ and $\omega$ imply

$$\kappa = \left|\frac{\partial \phi}{\partial x}\right| = \left|\frac{d\phi_1}{dx}\right|, or, \frac{d\phi_1}{dx} = \pm\kappa \tag{5}$$

and

$$\omega = \left|\frac{\partial \phi}{\partial t}\right| = \left|\frac{d\phi_2}{dt}\right|, or, \frac{d\phi_2}{dt} = \pm\omega \tag{5}$$

since
$$\phi(x,t) = \phi_1(x) + \phi_2(t)$$
.

There are two possibilities: $\frac{d\phi_1(x)}{dx}$ and $\frac{d\phi_2(t)}{dt}$ may be constants, or, these may be functions of x and t respectively.

We first consider the case when $\kappa$ and $\omega$ are constants. For a photon traveling along +x direction, the obvious choice is

$$\frac{d\phi_1}{dx} = \kappa, and, \frac{d\phi_2}{dt} = -\omega \tag{7}$$

which is just one of four alternative choices. Inserting the ansatz, Eq.(3), in Eq.(2), one obtains (provided $\psi \neq 0$)

$$\frac{1}{R_2}\frac{d^2 R_2}{dt^2} - \frac{2i\omega}{R_2}\frac{dR_2}{dt} - \omega^2 = c^2\left[\frac{1}{R_1}\frac{d^2 R_1}{dx^2} + \frac{2i\kappa}{R_1}\frac{dR_1}{dx} - \kappa^2\right]$$

$$= \text{a complex constant} = a_1 + ib_1, say, \tag{8}$$

where $a_1$ and $b_1$ are real. Equate the real parts of both sides of the first and last parts of Eq.(8) to obtain

$$\frac{1}{R_2}\frac{d^2 R_2}{dt^2} = \omega^2 + a_1 \tag{8a}$$

which yields

$$R_2(t) = A\exp[\pm(a_1 + \omega^2)^{\frac{1}{2}}t] \tag{9}$$

with a real constant A. Equating the imaginary parts of both sides of the first and last parts of Eq.(8), one obtains

$$\frac{2\omega}{R_2}\frac{dR_2}{dt} = -b_1.$$

This yields

$$R_2(t) = B\exp\left[\frac{b_1 t}{2\omega}\right] \tag{10}$$

where $B$ is a real constant. At $t = 0$, Eqs. (9) and (10) give

$$R_2(0) = A = B.$$

Equating the right sides of Eqs. (9) and (10), one has

$$\pm\sqrt{a_1 + \omega^2} = -\frac{b_1}{2\omega}. \tag{11}$$

Now equate the real parts of the second equation of Eq.(8) to find

$$\frac{c^2}{R_1}\frac{d^2 R_1}{dx^2} = a_1 + \omega^2. \tag{12}$$

The solution is

$$R_1(x) = B'\exp\left[\pm x\sqrt{\frac{a_1 + \omega^2}{c^2}}\right] \tag{13}$$

where $B'$ is a real constant. Similarly, equating the imaginary parts of both sides of the second equation of Eq.(8), one arrives at

$$\frac{2c^2\kappa}{R_1}\frac{dR_1}{dx} = b_1 \tag{13a}$$

which in turn shows that

$$R_1(x) = A'\exp\left[\frac{b_1 x}{2\omega c}\right]. \tag{14}$$

$A'$ is a real constant. At $x = 0$, Eqs.(13) and (14) gives

$$A' = B' = R_1(0).$$

When a photon approaches a measurement point (MP) at x, $P_1(x)$ will obviously decrease with increasing x [see the probability wave diagrams in Ref.37]

$$\frac{dP_1}{dx} = \frac{d}{dx}(R_1^2) = 2R_1 \frac{dR_1}{dx} < 0.$$

Since $R_1(x) = \sqrt{P_1(x)} > 0$, $\frac{dR_1}{dx}$ will be negative. Eq.(13a) then requires

$$b_1 < 0.$$

Let $b_1 = -b_1'$, where $b_1' > 0$. From Eqs.(14) and (10) respectively, the amplitudes become

$$R_1(x) = A' \exp\left[-\frac{b_1' x}{2\omega c}\right] \qquad (15a)$$

and

$$R_2(t) = B \exp\left[\frac{b_1' t}{2\omega}\right] \qquad (15b)$$

The above two equations combine to form the photon probability density for an incoming photon [ Fig.(1)] :

$$P(x,t) = R_1^2 R_2^2 = P(0,0) \exp\left[-\frac{b_1'}{\omega c}(x-ct)\right], \text{ for } x \geq ct \qquad (16)$$

When a photon recedes the MP (The point x is called the MP because $\psi(x,t)$ or $P(x,t)$ is measured at this point at a distance x from the origin) $P_1(x)$ would increase with increasing x, so

$$\frac{dP_1}{dx} = 2R_1 \frac{dR_1}{dx} > 0$$

implies $\frac{dR_1}{dx} > 0$. Eq.(13a), as also Eq.(10) require $b_1 > 0$. The result for a receding photon is [Fig.(2)]

$$P(x,t) = P(0,0) \exp\left[\frac{b_1}{\omega c}(x-ct)\right], \quad \text{for } x \leq ct. \qquad (16a)$$

Eqs. (16) and (16a) show that probability density is a wave traveling with velocity c (equal to photon speed) along +x direction. The dynamics of the probability wave has been described in detail in the context of quantum measurement problem in Ref.[37]. Time is measured such that when the photon is at x = 0 , t = 0. Hence, P(0,0) = 0. Referring to Fig.(1), where the photon is at a distance ct from the origin at time t, we normalize P(x,t) :

$$1 = \int_{ct}^{\infty} P(x,t)dx = \int_{ct}^{\infty} \exp\left[-\frac{b_1'}{\omega c}(x-ct)\right]dx .$$

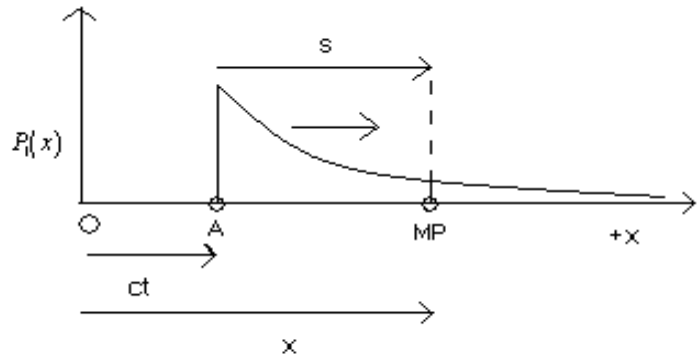

Fig.(1): A photon with a probability density wave peak at A, distant ct
The origin O, approaches the MP at x

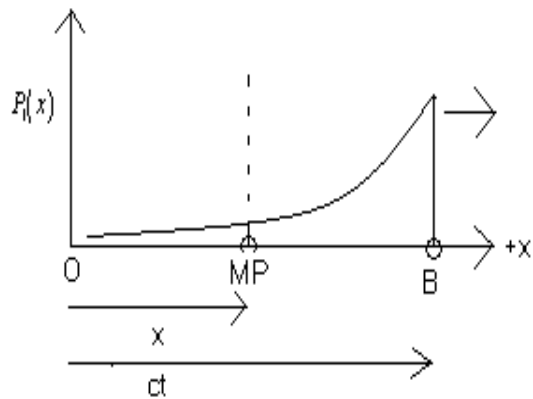

Fig.(2): A photon with a probability density wave peak at B, distant ct
from origin O, recedes the MP at x

Integration yields the value

$$b_1' = \omega c.$$

From Eq.(16), the probability density is

$$P(x,t) = \exp[-(x-ct)], \quad \text{for} \quad x \geq ct. \tag{17}$$

If position of the MP is measured from the photon [Fig.(1)], then

$$s = (x - ct) \geq 0$$

and probability of finding the photon at a distance $s$ from it may be found from Eq.(17):

$$P(s) = \exp(-s), \quad \text{for} \quad s \geq 0. \tag{18}$$

It is straightforward to show that

$$P(x,t) = \exp(x - ct), \quad \text{for} \quad x \leq ct \tag{19}$$

The fact that a photon is quantum-mechanically a probability wave (as the above equation says) rules out the objections to a photon wave function[42A,43,44]. While a conventional wave function of photon, e.g. Landau-Peierls function[45], consider it as having electromagnetic origin, we here explore only the ontological basis of a photon through its probabilistic origin. While photons are quanta of electromagnetic field, they are also quanta of probability field $P(s)$, described by Eq.(18). This equation also supports the observation of I.Biyalinicki-Birula [46] that photo-detection probability falls off exponentially.

There are mainly two objections to photon wave function :
(i) No position operator exists for photon,
(ii) While the position space wave function may be localized near a space-time point, the measurable quantities, like the electromagnetic field vectors, energy and photo-detection probability remain spread out.

To refute the objections, I first derive the photon wave function. Eq.(7) implies

$$\phi_1(x) = \kappa x,$$
$$\text{and} \quad \phi_2(t) = -\omega t$$

where I have ignored the arbitrary constants of integration. Making use of these and Eqs.(3),(17) and (19) the photon wave functions in the two domains take the following forms :

$$\psi(x,t) = \exp\left[-\frac{1}{2}(x-ct)+i\kappa x - i\omega t\right], \quad \text{for} \quad x \geq ct \qquad (20)$$

and
$$\psi(x,t) = \exp\left[-\frac{1}{2}(ct-x)+i\kappa x - i\omega t\right], \quad \text{for} \quad x \leq ct \qquad (21)$$

At the MP, viz. x = ct, the wave function becomes a stationary wave

$$\psi\left(x,\frac{x}{c}\right) = \exp(i\kappa x - i\omega t) \qquad (22)$$

so in the measurement event at spacetime point (x = ct, t) the probability density is

$$P\left(x,\frac{x}{c}\right) = 1,$$

implying that a whole photon has been found at x at time t = x/c. The MP characterized by x = ct is the one-dimensional analog of the three-dimensional MP

$$r = \sqrt{x^2 + y^2 + z^2} = ct.$$

It is interesting to note that measurement event is recorded at null interval :

$$c^2 t^2 - x^2 - y^2 - z^2 = 0.$$

The intervals shown in Eqs.(20) and (21) are respectively space-like and time-like. Quantizing the classical relation of position $s_p$ of a photon at time $t_0$ ,

$$s_p = ct_0$$

one obtains

$$\hat{s}_p \psi = ct_0 \psi . \qquad (23)$$

For a photon, $c = E/p$. Symmetrizing it and then inserting in Eq.(23), I obtain the eigenvalue equation after quantization:

$$\hat{s}_p \psi = \frac{1}{2}\left(\hat{E}\hat{p}^{-1} + \hat{p}^{-1}\hat{E}\right)t_0 \psi \qquad (24)$$

To be valid, the above equation requires the existence of invertibility of $\hat{p}$. To investigate this, we find the norm

$$\|\hat{p}\psi\| = \sqrt{\left\langle -i\hbar \frac{\partial \psi}{\partial x} \middle| -i\hbar \frac{\partial \psi}{\partial x} \right\rangle} = \hbar\sqrt{\kappa^2 + \frac{1}{4}} \|\psi\|$$

where $\psi$ is described by Eq.(20). If b is a positive number such that

$$\hbar\sqrt{\kappa^2 + \frac{1}{4}} \geq b,$$

then $\|\hat{p}\psi\| \geq b\|\psi\|$ for every $\psi \in dom(\hat{p})$. Therefore $\hat{p}$ admits a continuous inverse $\hat{p}^{-1}$. To find its eigenvalue, note that

$$\hat{p}^{-1}\hat{p}\psi = \psi,$$

or, using the explicit form of $\psi$ given in Eq.(20), the above equation results in

$$\hat{p}^{-1}\psi = \left[ \frac{1}{\hbar\kappa + \frac{1}{2}i\hbar} \right]\psi. \tag{25}$$

The required eigenvalue of $\hat{p}^{-1}$ is $\left[ \dfrac{1}{\hbar\kappa + \frac{1}{2}i\hbar} \right].$

We are now in a position to study Eq.(24). Using Eq.(25), Eq.(24) gives

$$\hat{s}_p\psi = \frac{t_0}{2}\left[ \frac{\hat{E}\psi}{\hbar\kappa + \frac{1}{2}i\hbar} + \hat{p}^{-1}\left(i\hbar \frac{\partial \psi}{\partial t}\right)\right] = \frac{i\hbar t_0\left(\frac{c}{2} - i\omega\right)}{2} \frac{2\psi}{\left(\hbar\kappa + \frac{1}{2}i\hbar\right)} = ct_0\psi \tag{26}$$

where, as usual, $\psi$ is described by Eq.(20). This eigenvalue equation unambiguously proves that the position operator of a photon exists and its explicit form at time $t_0$ is

$$\hat{s}_p = \frac{t_0}{2}\left( i\hbar \frac{\partial}{\partial t}\hat{p}^{-1} + \hat{p}^{-1}i\hbar\frac{\partial}{\partial t}\right). \tag{27}$$

while its eigenvalue at that time is simply $ct_0$. The eigenvalue is real. This sets apart photons from all other particles in Quantum Mechanics. While I have shown elsewhere [37] that all quantum systems travel in complex spacetime except at the measurement event, photons always move in real spacetime. This might provide a reason why photons have the maximum permissible speed in real-dimensional nature!

The other objection to a photon wave function may be refuted in quantum-mechanical terms. It is reiterated that the photon wave function derived here has no relation whatsoever with the wave function derived from electromagnetic inputs. Here, the photon is a probability wave ------ a generic mathematical wave in probability space. Indeed all quantum systems are probability waves [37] when left to evolve i.e. when they are not subjected to measurements. The probability fields of different quantum particles do not interact among themselves because these are mathematical (probability) fields. The observables of a photon may be calculated using the appropriate operators on the photon wave functions, Eqs.(20) and (21). Of course, the corresponding eigenvalues (except position) may be complex. This only shows that the corresponding operators are non-Hermitian normal operators which transform into Hermitian ones at the MP [37]. At the MP the measurement device will record real numbers ---- so there is no problem. Let us see that it is really so. In general, when the photon is not at the MP, i.e., $x \neq ct$,

$$\hat{E}\psi = i\hbar \frac{\partial \psi}{\partial t} = \left(\hbar\omega + \frac{1}{2}i\hbar c\right)\psi .$$

But at the MP, x = ct, and $\psi = \exp(i\kappa x - i\omega t)$, and so

$$\hat{E}\psi = \hbar\omega\psi .$$

This reveals the transformation of non-Hermitian operators into Hermitian ones at the MP. The complex eigenenergies need not be cofused with those characteristic of decay or growing Gamow vectors. These are mere consequences of the complex spacetime manifold in which the quantum particles live.

So far I have discussed the case when $\frac{d\phi_1}{dx}$ and $\frac{d\phi_2}{dt}$ are constants. Now I consider the fact that these derivatives need not mean that they are not functions of x and t respectively. This possibility introduces tachyons in complex spacetime.

# 3. WHEN $\frac{d\phi_1(x)}{dx}$ AND $\frac{d\phi_2(t)}{dt}$ ARE NOT CONSTANTS

I therefore rewrite the wave number and frequency as

$$\frac{d\phi_1(x)}{dx} = \pm \kappa(x), \text{ and } \frac{d\phi_2(t)}{dt} = \pm \omega(t)$$

and for a photon traveling in +x direction, I choose

$$\frac{d\phi_1(x)}{dx} = \kappa(x) \text{ and } \frac{d\phi_2(t)}{dt} = -\omega(t)$$

Since $\frac{\omega(t)}{\kappa(x)} = c(x,t) = c_1(x)c_2(t)$, where separation of variables have been assumed, one finds

$$\kappa(x) = \frac{1}{c_1(x)} \text{ and } \omega(t) = c_2(t).$$

Keeping these in mind, I insert Eq.(3) in Eq.(2) and find

$$\frac{1}{R_2}\frac{d^2 R_2}{dt^2} - \frac{2i\omega}{R_2}\frac{dR_2}{dt} - i\frac{d\omega}{dt} - \omega^2 = c^2 \left[ \frac{1}{R_1}\frac{d^2 R_1}{dx^2} + \frac{2i\kappa}{R_1}\frac{dR_1}{dx} + i\frac{d\kappa}{dx} - \kappa^2 \right] \tag{28}$$

The above equation generates four equations when real and imaginary parts of both sides are equated :

$$\frac{1}{c_2^2}\left[\frac{1}{R_2}\frac{d^2 R_2}{dt^2} - \omega^2\right] = c_1^2\left[\frac{1}{R_1}\frac{d^2 R_1}{dx^2} - \kappa^2\right] = \text{a real constant} = a, \tag{29a}$$

$$\left[-\frac{1}{c_2^2}\right]\left[\frac{d\omega}{dt} + \frac{2\omega}{R_2}\frac{dR_2}{dt}\right] = c_1^2\left[\frac{d\kappa}{dx} + \frac{2\kappa}{R_1}\frac{dR_1}{dx}\right] = \text{a real constant} = b. \tag{29b}$$

Substituting $[1/\kappa(x)]$ and $\omega(t)$ for $c_1(x)$ and $c_2(t)$ respectively the following equations are obtained :

$$\frac{1}{R_2}\frac{d^2 R_2}{dt^2} - \omega^2 = a\omega^2 \tag{30a}$$

$$\frac{1}{R_1}\frac{d^2 R_1}{dx^2} - \kappa^2 = a\kappa^2 \tag{30b}$$

$$\frac{d\omega}{dt} + \frac{2\omega}{R_2}\frac{dR_2}{dt} = -b\omega^2 \tag{30c}$$

$$\frac{d\kappa}{dx} + \frac{2\kappa}{R_1}\frac{dR_1}{dx} = b\kappa^2 \tag{30d}$$

From Eq.(30c), a little rearrangement gives

$$\frac{d\omega}{\omega} + \frac{2dR_2}{R_2} = -b\omega dt$$

which after integration results into

$$R_2(t) = \sqrt{\frac{A}{\omega}}\exp\left[-\frac{b}{2}\int \omega dt\right] \tag{30}$$

where A is a real constant. Eq.(30d) is similar to Eq.(30c) except for the sign on the right side. Hence its solution may be immediately written down :

$$R_1(x) = \sqrt{\frac{B}{\kappa}}\exp\left[\frac{b}{2}\int \kappa dx\right] \tag{31a}$$

Since $[\phi_1(x) + \phi_2(t)] = (\int \kappa dx - \int \omega dt)$, Eqs.(30) and (31a) help write down the photon wave function from Eq.(3) :

$$\psi(x,t) = \sqrt{\frac{AB}{\omega\kappa}}\exp\left[\frac{b}{2}\int \kappa dx - \frac{b}{2}\int \omega dt + i\int \kappa dx - i\int \omega dt\right] \tag{31b}$$

and

$$P(x,t) = \left[\frac{D}{\omega\kappa}\right]\exp\left[b\left(\int \kappa dx - \int \omega dt\right)\right] \tag{31b}$$

where AB = D and $x \geq \int c(x,t)dt$. From Eq.(31b), P(0,0) = 1 means
D = $\omega(0)\kappa(0) = \omega_0\kappa_0$. Hence

$$P(x,t) = \frac{\omega_0\kappa_0}{\omega\kappa}\exp\left[b(\int \kappa dx - \int \omega dt)\right] \tag{31c}$$

But $\int \kappa dx - \int \omega dt = \int \kappa(x)[dx - c(x,t)dt]$, and $x \geq \int c(x,t)dt$ implies that

$$[dx - c(x,t)dt] \geq 0, \text{ so,} \quad P(x,t) = \frac{\omega_0\kappa_0}{\omega\kappa}\exp\left[\int b\kappa\{dx - cdt\}\right] . \tag{32}$$

$\kappa(x)$ is positive as the photon is traveling along the +x direction. Now a look at a similar photon wave function of constant speed, viz. Eq.(20) for $x \geq ct$ suggests that b in the exponent of Eq. (32) must be negative in order to check the diverging P(x ,t). We set $b = -b'$, where $b'$ is positive. Eq. (32) then describes probability density as an inhomogeneous wave traveling with a varying speed c(x ,t) along +x direction. The photon wave function may now be written down from Eq. (31b):

$$\psi(x,t) = \sqrt{\frac{\omega_0 \kappa_0}{\omega \kappa}} \exp\left[\frac{\int b' c_1 c_2 dt}{2 c_1} - \frac{b'}{2} \int \frac{dx}{c_1} + i \int \kappa dx - i \int \omega dt\right] \quad (32a)$$

or,

$$\psi(x,t) = \sqrt{\frac{\omega_0 \kappa_0}{\omega \kappa}} \exp\left[-\frac{b'}{2}\left\{\int \kappa dx - \int \omega dt\right\} + i \int \kappa dx - i \int \omega dt\right] \quad (32b)$$

where $x \geq \int c(x,t) dt$.

The rule for finding the value R of an observable $\hat{R}$ in a particular measurement is [37] :

$$R = \frac{\hat{R}\psi}{\psi},$$

and therefore, the photon energy in a particular measurement , calculated from Eq. (32b), is

$$E_r = \frac{\hat{E}\psi}{\psi} = \hbar\omega(t) + \frac{1}{2} i\hbar b' \omega(t) - \frac{i\hbar}{2\omega}\frac{d\omega}{dt} \quad (33)$$

The photon momentum may be similarly found :

$$p_r = \frac{\hat{p}\psi}{\psi} = \hbar\kappa(x) + \frac{1}{2} i\hbar b' \kappa(x) + \frac{i\hbar}{2\kappa}\frac{d\kappa}{dx} \quad (34)$$

The energy and momentum values of a photon are related as

$$E_r = p_r c(x,t) = p_r c_1(x) c_2(t) \quad (34A)$$

Inserting the values of $E_r$ and $p_r$ from Eqs.(33) and (34) respectively, in Eq.(34A) and replacing $\omega(t)$ and $\kappa(x)$ by $c_2(t)$ and $\frac{1}{c_1(x)}$ one arrives at

$$\hbar\omega(t) + \frac{1}{2}i\hbar b'\omega(t) - \frac{i\hbar}{2\omega}\frac{d\omega}{dt} = c_1 c_2 \left[ \hbar\kappa(x) + \frac{1}{2}i\hbar b'\kappa(x) + \frac{i\hbar}{2\kappa}\frac{d\kappa}{dx} \right] \tag{34B}$$

which is separated into real and imaginary parts and a little rearrangement leads to the important result :

$$\frac{1}{c_2^2}\frac{dc_2}{dt} = \frac{dc_1}{dx} = \lambda \tag{34a}$$

where $\lambda$ is a real constant. The above two equations yield the following solutions :

$$c_1(x) = c_1(0) + \lambda x \tag{35}$$

$$c_2(t) = \frac{c_2(0)}{1 - \lambda t c_2(0)} \tag{36}$$

The space-time varying speed becomes

$$c(x,t) = \frac{c_2(0)[c_1(0) + \lambda x]}{1 - \lambda t c_2(0)} \tag{36A}$$

From Eq. (34a) one finds

$$\frac{\partial c}{\partial t} = c_1 \frac{dc_2}{dt} = \lambda c_1 c_2^2$$

$$c\frac{\partial c}{\partial x} = c_1 c_2^{\ 2} \frac{dc_1}{dx} = \lambda c_1 c_2^2,$$

and the above two equations easily presents a differential equation for varying speed of light :

$$\frac{\partial c}{\partial t} = c\frac{\partial c}{\partial x} \tag{36a}$$

Varying speed of light (VSL) theories [47,48,49,50,51] have always received fatal blows while trying to interpret the varying fine structure constant results. While interpreting the varying fine structure constant results the lethal line is that all lengths, masses and times are measured in dimensionless numbers (multiples) of unit length, mass and time. A change in a dimensionful quantity cannot rule out the possibility that the meter rods ,clocks etc have undergone similar changes in lengths ticks etc. No experiment can distinguish ! Therefore change in a dimensionful quantity does not point towards a change in a dimensionless quantity like $\alpha$.

But all these arguments hold only when we measure something like speed of light with `meter` of a meter ruler and `second` of a clock, which are all real numbers.That is , the measurements involve four-dimensional real spacetime. But here, we find that something extraordinary happens. From Eq. (34A), the varying speed of light is

$$c(x,t) = \frac{E_r}{p_r} = \frac{\hbar\omega(t) + \frac{1}{2}i\hbar b'\omega(t) - \frac{i\hbar}{2\omega}\frac{d\omega}{dt}}{\hbar\kappa(x) + \frac{1}{2}ib'\kappa(x) + \frac{i\hbar}{2\kappa}\frac{d\kappa}{dx}} \tag{36b}$$

where we have used Eqs.(33) and (34). This speed c(x ,t ) is a compex quantity, and so it cannot be measured by real meters and seconds. In fact these speed is not measurable. Whenever we measure speed of light this space-time varying complex speed transforms into a real valued constant speed $c(0,0) = c_0$, which is the present speed of light in vacuum. To prove how this measured value of a complex c(x ,t ) transforms into a real c(0,0) we recall that the criterion of measurement is spelt out as

$$x = \int c(x,t)dt \tag{36c}$$

from which we find

$$\frac{x}{c_1(x)} = \int_0^t c_2(t)dt = \text{a real constant} \tag{36d}$$

Differential of the second equation gives $c_2(t)dt = 0$. Since $c_2(t)$ cannot be zero (otherwise c(x ,t ) would be zero), $dt = 0$, i.e., $t = t_0$ = a constant at the measurement event. Since dt = 0, differentiating Eq.(36c) yields

$$dx = c(x,t)dt = 0, \text{ i.e., } x = x_0 = \text{a constant.}$$

If we want to measure c(x ,t ) at the MP, then we replace x and t in Eq. (36b) by $x_0$ and $t_0$. Now we use the relation $\omega(t_0) = \kappa(x_0)c(x_0,t_0)$ in Eq. (36b) to obtain

$$c(x,t) = c(x_0,t_0) \tag{36e}$$

If we set the present time as t = 0 = $t_0$, then Eq.(36d) gives x = 0. Hence from Eq.(36e) the speed of light at the MP is $c(0,0) = c_0$.

   Gravitational redshift of light rising away from a distant quaser is caused by a decrease in frequency :

$$\frac{d\omega}{dt} < 0, \text{ i.e., } \frac{dc_2}{dt} < 0.$$

Eq.(34a) now says $\quad \lambda = -\lambda_0 < 0.$

If we assume the present spacetime as origin (t = 0, x = 0) located in the earth-bound laboratory, then the look-back time $t = -T < 0$, and the look-back distance of the quaser,(- x) constitutes the measurement point. Eqs.(35) and (36) change, but are not altered in form

$$c_1(x) = c_1(0) + \lambda_0 x,$$

$$c_2(T) = \frac{c_2(0)}{1 - \lambda_0 c_2(0) T}$$

From above, $c_1(x) > c_1(0)$ and $c_2(T) > c_2(0)$,
whence $c(x,T) = c_1(x) c_2(T) > c_1(0) c_2(0) = c_0$
The varying speed of light is then

$$c(x,T) = \left[ \frac{\{c_1(0) + \lambda_0 x\} c_2(0)}{1 - \lambda_0 c_2(0) T} \right]$$

Since the above equation and Eq.(36A) are similar in all respects we shall drop $\lambda_0$ and T in future discussion, and instead continue to use $\lambda$ and t and Eq.(36A).
I have thus proved that c ( x t ) > $c_0$, an astonishing result that says, older photons had higher speed than $c_0$. Special theory of relativity categorically says those older particles were certainly not photons. But tachyons [52,53] may be suitable candidates satisfying this faster-than-light speed. Note that

$$c(x,t) = c_1 c_2 = c_1(x) \omega(t) = \frac{c_1 E(t)}{\hbar}$$

and this implies an energy-dependence of speed of light; quantum gravity models suggest just this [54,55].
A few difficulties are now to be resolved once I propose tachyons[56]. An infinite amount of energy is required to slow a tachyon down to the speed of light. But theory requiring this is the classical special relativity. In Quantum theory I have proved through Eqs. (36c),(36d) and (36e) how c ( x t ) , a higher speed, transforms into the lower speed $c_0$ at the MP.
The violation of causality by tachyons never takes place because emission of a tachyon at a measurement event E and its absorption at event E' actually involve photons, not tachyons. This is due to the fact that tachyons show up as photons at the measurement events. This is possible because the least possible speed of tachyon is speed of light. The events E and E' involve photons ,so causality is not flouted. As tachyons move in complex spacetime they will forever remain unobservable. The 26-dimensional string theory requires tachyons.The pernicious negative norm states are however no longer dreaded objects as the latter have been shown to be perfectly compatible with Quantum Mechanics [37]. Superstring theory has replaced bosonic string theory because the latter suffered from tachyonic instability. But it will become clear later that tachyons in complex spacetime having real (and not imaginary) mass, with, of course, imaginary energy and momentum (not surprising in complex manifold) may not be a disaster.

For tachyons with $v > c$, energy and momentum

$$p = \frac{imv}{\sqrt{\frac{v^2}{c^2} - 1}}, \tag{37A}$$

$$E = \frac{imc^2}{\sqrt{\frac{v^2}{c^2} - 1}} \tag{37B}$$

are imaginary quantities. This is not anarchy ---- as tachyons move exclusively in complex spacetime and never show up at real spacetime. I do not attribute the peculiar tachyon properties in the conventional way: assigning imaginary invariant mass to tachyons. Rather I keep tachyon mass real while E and p are imaginary even in the classical domain, as tachyons have never been detected at classical spacetime points at a measurement event.

Quantizing Eq.(37B)

$$\hat{E}\psi = i\hbar \frac{\partial \psi}{\partial t} = \frac{imc^2 \psi}{\sqrt{\frac{\hat{v}^2}{c^2} - 1}} = i\hat{m}_r c^2 \psi \tag{37}$$

where $\hat{v}$ is assumed a multiplicative operator (as it works!) and $\hat{m}_r$ is the relativistic mass operator. Explicit forms of two operators are in order : From Eq.(37), the operator

$$\hat{\gamma} = \frac{1}{\sqrt{\frac{\hat{v}^2}{c^2} - 1}} = \frac{\hbar}{mc^2} \frac{\partial}{\partial t} \tag{38}$$

which may be of immense help in quantizing important classical relations of special relativity. From the same equation, the relativistic mass operator is found:

$$\hat{m}_r = \frac{m}{\sqrt{\frac{\hat{v}^2}{c^2} - 1}} = \frac{\hbar}{c^2} \frac{\partial}{\partial t} \tag{39}$$

To find the tachyon wave function, I insert the ansatz

$$\psi(x,t) = \psi_1(x) R_2(t) \exp[i\phi_2(t)]$$

into

$$i\hbar \frac{\partial \psi}{\partial t} = i m_r c^2 \psi$$

to obtain

$$\frac{i\hbar}{R_2} \frac{dR_2}{dt} + \hbar\omega = i m_r c^2 .$$

This gives the expected result: The real part of tachyon energy is zero, $\hbar\omega = 0$. The imaginary parts of both sides, when equated and solved, yield

$$R_2(t) = A \exp\left[\int \frac{m_r c^2}{\hbar} dt\right] \qquad (41)$$

where A is a real constant. Making use of

$$m_r = \frac{mc}{\sqrt{v^2 - c^2}}$$

one gets

$$R_2(t) = A \exp\left[\frac{mc^3}{\hbar} \int \frac{dt}{\sqrt{v^2 - c^2}}\right] \qquad (42)$$

Tachyon is a massive particle and its invariant mass is m. It is quite natural now to find the least possible mass allowed in Quantum Mechanics in order to fit the roles of both photons and tachyons. Photons in classical special relativity are massless but their quantum-mechanical counterparts in complex spacetime (tachyons) are massive. Since tachyons appear as photons at the MP, photons are massive too, since mass cannot simply be wished away. I now explore Klein-Gordon equation for this least mass of scalar tachyons, and on the way try to refute the objections that have made its use questionable time and again as a single-particle equation.

## 4. KLEIN-GORDON EQUATION AND QUANTUM-MECHANICAL ANALOG OF THE CLASSICAL FORMULA $E = MC^2$

We rewrite Klein-Gordon equation as

$$\frac{1}{c^2} \frac{\partial^2 \psi}{\partial t^2} - \frac{\partial^2 \psi}{\partial x^2} + \kappa_c^2 \psi = 0 \qquad (43)$$

where $\kappa_c = \frac{mc}{\hbar}$. Inserting the ansatz of Eq.(3) into Eq.(43), one obtains

$$\frac{1}{c^2}\left[\frac{1}{R_2}\frac{d^2R_2}{dt^2} - \frac{2i\omega}{R_2}\frac{dR_2}{dt} - \omega^2\right] - \left[\frac{1}{R_1}\frac{d^2R_1}{dx^2} + \frac{2i\kappa}{R_1}\frac{dR_1}{dx} - \kappa^2\right] + \kappa_c^2 = 0$$

Equating the real parts of both sides,

$$\frac{1}{c^2 R_2}\frac{d^2R_2}{dt^2} - \frac{\omega^2}{c^2} + \kappa_c^2 = \frac{1}{R_1}\frac{d^2R_1}{dx^2} - \kappa^2 = \alpha \tag{44}$$

where $\alpha$ is a real constant. Equating the imaginary parts of both sides,

$$-\frac{2\omega}{R_2 c^2}\frac{dR_2}{dt} = \frac{2\kappa}{R_1}\frac{dR_1}{dx} = -\beta, \tag{45}$$

where $\beta$ is a real constant. Eq.(45) leads to

$$R_2(t) = A'\exp\left[\frac{\beta c^2 t}{2\omega}\right] \tag{46}$$

where $A'$ is a real constant. The second equation of Eq.(45) yields

$$R_1(x) = B'\exp\left[-\frac{\beta x}{2\kappa}\right] \tag{47}$$

$B'$ being a real constant. Definitions of wave number $\kappa$ and frequency $\omega$, as discussed earlier, imply

$$\frac{d\phi_1}{dx} = \pm\kappa; \quad \text{and,} \quad \frac{d\phi_2}{dt} = \pm\omega.$$

There are four possibilities for wave function of a Klein-Gordon particle:

$$\frac{d\phi_1}{dx} = +\kappa, \frac{d\phi_2}{dt} = +\omega$$

$$\frac{d\phi_1}{dx} = -\kappa, \frac{d\phi_2}{dt} = -\omega$$

$$\frac{d\phi_1}{dx} = -\kappa, \frac{d\phi_2}{dt} = +\omega$$

$$\frac{d\phi_1}{dx} = +\kappa, \frac{d\phi_2}{dt} = -\omega.$$

The solution of Klein-Gordon equation is then a four-component wave-function :

$$\psi(x,t) = \begin{bmatrix} \psi_{+\kappa,+\omega} \\ \psi_{-\kappa,-\omega} \\ \psi_{-\kappa,+\omega} \\ \psi_{+\kappa,-\omega} \end{bmatrix}$$

resembling wave function of Dirac equation. From Eqs.(46) and (47) the Klein-Gordon wave function may be written as

$$\psi(x,t) = A'B' \exp\left[\frac{\beta c^2 t}{2\omega} - \frac{\beta x}{2\kappa} + i\kappa x - i\omega t\right] \tag{48}$$

where $\kappa$ and $\omega$ are respectively $\dfrac{d\phi_1}{dx}$ and $-\dfrac{d\phi_2}{dt}$. Eq.(48) describes a free particle with positive energy moving along + x direction. The eigenvalues of observables are complex , and the non-Hermitian normal observables become Hermitian at the MP. The wave functions are defined in Rigged Hilbert space.
When the particle is at x = 0 at t = 0, P(0,0) = 1. Eq.(48) now reads

$$\psi(0,0) = A'B' = \sqrt{P(0,0)} = 1.$$

Hence, the wave function is purely exponential :

$$\psi(x,t) = \exp\left[\frac{\beta}{2}\left(\frac{c^2 t}{\omega} - \frac{x}{\kappa}\right) + i\kappa x - i\omega t\right] \tag{49}$$

and

$$P(x,t) = \exp\left[\beta\left(\frac{c^2 t}{\omega} - \frac{x}{\kappa}\right)\right].$$

Since the quantum system is situated at vt [cf. Fig.(1)], normalization yields

$$1 = \int_{vt}^{\infty} P(x,t)dx = \int_{vt}^{\infty} \exp\left[\beta\left(\frac{c^2 t}{\omega} - \frac{x}{\kappa}\right)\right]dx$$

or, $$1 = \frac{\kappa}{\beta}\exp\left[\frac{\beta c^2 t}{\omega} - \frac{\beta vt}{\kappa}\right]. \tag{49A}$$

Since the left side is independent of t , one gets

$$\frac{c^2}{\omega} = \frac{v}{\kappa} \tag{49B}$$

since $\beta$ is not zero, i.e. $R_1(x)$, $R_2(t)$ and therefore $P(x,t)$ are not constants.
Eq.(49A) now requires $\kappa = \beta$. (49C)

Inserting these results into Eq.(49) the Klein-Gordon wave function reads

$$\psi(x,t) = \exp\left[\frac{1}{2}(vt - x) + i\kappa x - i\omega t\right] \quad \text{for} \quad x \geq vt \tag{50}$$

whence the probability density P(x,t) is a wave traveling with particle speed v along +x direction :

$$P(x,t) = \exp[-(x - vt)] \quad \text{for} \quad x \geq vt$$
or, $$P(s) = \exp(-s) \tag{50A}$$

where s = (x -vt) is the separation between the position of the quantum system and MP. When s = 0, the two reference frames of the quantum system and the MP (device) particle (situated at their respective origins) coincide and a measurement event occurs at the specific spacetime point (x , x/v ). Eq. (50A) shows that P(0) = 1, and from Eq.(50),

$$\psi(x = vt) = \exp(i\kappa x - i\omega t) \tag{51}$$

and all the observables acting on Eq.(51) generate real eigenvalues at the MP. It has been proved [37] that this space-time point ( x , x/v ) is in the classical domain. The stationary waves extracted from the wave functions $\psi_{+\kappa,+\omega}$ and $\psi_{-\kappa,+\omega}$ at this classical point produce negative energies, and therefore, these may be dismissed as unphysical, as is usually done in classical physics. But this dismissal is legitimate only at the MP. When $x \neq vt$ the non-stationary solutions are perfectly valid since they yield complex eigenenergies.

The probability density for a Klein-Gordon particle is usually written as

$$P(x,t) = \frac{i\hbar}{2mc^2}\left[\psi^* \frac{\partial \psi}{\partial t} - \frac{\partial \psi^*}{\partial t}\psi\right] \tag{52}$$

The first objection to Klein-Gordon equation is that P(x ,t ) described by Eq.(52) is not positive definite. From Eq.(50) we pick up

$$\psi(x,t) = \exp\left[\frac{1}{2}(vt - x) + i\kappa x - i\omega t\right], \quad \text{for} \quad x \geq vt$$

so that
$$\frac{\partial \psi}{\partial t} = \left(\frac{1}{2}v - i\omega\right)\psi$$

and
$$\frac{\partial \psi^*}{\partial t} = \left(\frac{1}{2}v + i\omega\right)\psi^*.$$

Inserting these into Eq.(52) I find

$$P(x,t) = \frac{\hbar\omega}{mc^2}\exp[-(x-vt)]$$

which is clearly positive definite.

The second objection regarding negative energy solutions has already been addressed. The third objection circles around the superposition of real positive and negative energy solutions of the form

$$\psi(x,t) = a\psi_+ + b\psi_-$$

where $\psi_+$ and $\psi_-$ denote energy eigenfunctions of positive and negative eigenvalues. As we have already discarded the negative energy solutions as unphysical states at the classical spacetime point, the difficulty simply disappears.

Now we are ready to calculate the least possible invariant mass. From Eq. (47) I find

$$\frac{1}{R_1}\frac{d^2 R_1}{dx^2} = \frac{\beta^2}{4\kappa^2}$$

and from Eq.(44)

$$\frac{\beta^2}{4\kappa^2} - \kappa^2 = \alpha. \tag{54}$$

Eq.(46) yields

$$\frac{1}{R_2}\frac{d^2 R_2}{dt^2} = \frac{\beta^2 c^4}{4\omega^2},$$

and from Eq.(44), with the help of above results,

$$\frac{\beta^2 c^2}{4\omega^2} - \frac{\omega^2}{c^2} + \frac{m^2 c^2}{\hbar^2} = \alpha. \tag{55}$$

Eqs.(54) and (55) give

$$\frac{\beta^2 c^2}{4\omega^2} - \frac{\omega^2}{c^2} + \frac{m^2 c^2}{\hbar^2} = \frac{\beta^2}{4\kappa^2} - \kappa^2$$

which in turn becomes

$$\omega^2 = \kappa^2 c^2 + \frac{m^2 c^4}{\hbar^2} + \frac{\kappa^2 c^4}{4\omega^2} - \frac{c^2}{4}$$

because $\beta = \kappa$ [from Eq.(49C)]. And so,

$$E^2 = \hbar^2 \omega^2 = p^2 c^2 + m^2 c^4 + \frac{\hbar^2}{4}\left[\left(\frac{\kappa^2}{\omega^2}\right)c^4 - c^2\right]. \tag{55a}$$

Eq.(49B) gives

$$\frac{\kappa}{\omega} = \frac{v}{c^2}. \tag{55A}$$

Hence,
$$E^2 = p^2 c^2 + m^2 c^4 - \frac{\hbar^2 c^2}{4}\left(1 - \frac{v^2}{c^2}\right). \tag{56}$$

When the Klein-Gordon particle is at rest, p = 0 and v = 0. Eq.(56) now becomes

$$E^2 = m^2 c^4 - \frac{\hbar^2 c^2}{4}. \tag{57}$$

The lower bound for E for this free particle is zero. In such a case, Eq.(57) yields the least possible quantum-mechanical mass :

$$m = \frac{\hbar}{2c} = 1.7 \times 10^{-38} \, gm. \tag{57A}$$

I assign this least mass to the tachyon concerned ,which hands over this mass to photon at the measurement event. This value of photon invariant mass is exceptionally close to those obtained in Gamma Ray Bursts (GRB) events [77,78]

(i) $m = 1.7 \times 10^{-38} \, gm.$ ,event GRB 910607 (1998)

(ii) $m = 7.4 \times 10^{-38} \, gm.$ ,event GRB 930131(1994)

This non-zero photon rest mass would bring in trouble for Quantum Electrodynamics through loss of gauge invariance, making it non-renormalizable. Also, charge conservation would not be absolutely guaranteed. But all these should not discourage an experiment devised to test the predicted photon mass given by Eq.(57A) [79,80].

Since the momentum

$$p = \frac{mv}{\sqrt{1-\frac{v^2}{c^2}}} = Mv$$

Eq.(56) reads

$$E^2 = M^2 c^4 \left[1 - \frac{\hbar^2}{4M^2 c^2}\left(1 - \frac{v^2}{c^2}\right)\right]$$

which yields

$$E \cong Mc^2 - \frac{\hbar^2}{8M}\left(1 - \frac{v^2}{c^2}\right) \tag{58}$$

This is the quantum-mechanical analog of the classical equation

$$E = Mc^2.$$

The correction term in Eq. (58) is due to quantum potential ( or, as will become clear later, due to quantum gravitational potential arising out of particle mass). Eq. (58) may be tested experimentally to check its validity.

# 5. CALCULATION OF VARYING FINE STRUCTURE CONSTANT

With tachyon mass

$$m = \frac{\hbar}{2c},$$

Eq.(42) reads

$$R_2(t) = A \exp\left[\frac{c^2}{2}\int \frac{dt}{\sqrt{v^2 - c^2}}\right] \tag{59}$$

The tachyon speed $v = c(x,t) = c_1(x)c_2(t)$ while photon speed $c = c(0,0) = c_1(0)c_2(0)$. Without loss of generality, the MP may be chosen at $x = 0$, so that

$$v(0,t) = c(0,t) = c_1(0)c_2(t).$$

Therefore,

$$\frac{c^2}{2}\int \frac{dt}{v^2 - c^2} = \frac{1}{2}c_2^2(0)\int \left[\left(\frac{dc_2}{c_2^2 - c_2^2(0)}\right)\frac{1}{\left(\frac{dc_2}{dt}\right)}\right]$$

$$= \frac{1}{2}c_2^2(0)\int \left[\left(\frac{dc_2}{c_2^2 - c_2^2(0)}\right)\frac{1}{(\lambda c_2^2)}\right]$$

where I have used Eq.(34a). Inserting the above expression in Eq.(59) and assuming the change in variable

$$c_2(t) = c_2(0)\sec\theta,$$

we arrive at

$$R_2(t) = A\exp\left[\frac{1}{2\lambda}\sin\sec^{-1}\frac{c_2(t)}{c_2(0)}\right] \tag{60}$$

and

$$R_2(0) = A\exp\left[\frac{1}{2\lambda}\sin\sec^{-1}1\right] = A$$

Now,

$$\frac{dR_2}{dt} = \frac{dR_2}{dc_2}\left(\frac{dc_2}{dt}\right) = \lambda c_2^2(t)\left(\frac{dR_2}{dc_2}\right)$$

and from Eq.(60)

$$\left(\frac{dR_2}{dc_2}\right) = \frac{R_2(0)}{2\lambda}\frac{d}{dc_2}(\sin\theta)\exp\left[\frac{1}{2\lambda}\sin\theta\right]$$

where $\theta = \sec^{-1}\left[\frac{c_2(t)}{c_2(0)}\right]$. Therefore,

$$\frac{dR_2}{dt} = \frac{1}{2}R_2(0)c_2^2(t)\cos\theta \frac{d\theta}{dc_2}\exp\left[\frac{1}{2\lambda}\sin\theta\right].$$

But

$$\frac{d\theta}{dc_2} = \frac{1}{\left(\frac{dc_2}{d\theta}\right)} = \frac{1}{c_2(0)\sec\theta\tan\theta}.$$

Hence

$$\frac{dR_2}{dt} = \frac{c_2^2(t)}{2c_2(0)}\cot\theta\cos^2\theta R_2(t) \qquad (61)$$

where we have made use of Eq.(60). From Eq.(30c) with $\omega(t) = c_2(t)$,

$$\frac{2c_2(t)}{R_2}\frac{dR_2}{dt} + \frac{dc_2}{dt} = -bc_2^2(t) \qquad (62)$$

Since $\frac{dR_2}{dt}$ and $\frac{dc_2}{dt}$ are positive, $-b = b' > 0$. With help from Eq.(61), Eq.(62) becomes

$$\frac{c_2^3(t)}{c_2(0)}\cot\theta\cos^2\theta + \lambda c_2^2(t) = b'c_2^2(t). \qquad (63)$$

From Eq.(31b) with $D = \omega_0 \kappa_0$,

$$P(x,t) = \left(\frac{\omega_0 \kappa_0}{\omega \kappa}\right)\exp\left[b\int\frac{dx}{c_1(x)} - b\int c_2(t)dt\right] \qquad (63A)$$

Note that

$$\int\frac{dx}{c_1(x)} = \int\frac{dx}{c_1(0) + \lambda x} = \ln\left|c_1(0) + \lambda x\right|^{\left(\frac{1}{\lambda}\right)} \qquad (64)$$

and

$$\int c_2(t)dt = \int\frac{c_2(0)dt}{1 - \lambda c_2(0)t} = \ln\left|1 - \lambda c_2(0)t\right|^{\left(-\frac{1}{\lambda}\right)}. \qquad (65)$$

Utilizing the results of Eqs.(64) and (65) in Eq.(63A), one obtains

$$P(x,t) = \left[\frac{c_1(x)c_2(0)}{c_1(0)c_2(t)}\right]\exp\left[\ln|c_1(0)+\lambda x|^{\frac{b}{\lambda}} + \ln|1-\lambda c_2(0)t|^{\frac{b}{\lambda}}\right]$$

$$= \frac{1}{c_1(0)}[c_1(0)+\lambda x][1-\lambda c_2(0)t]\left|\lambda(x-c_0t)+c_1(0)-\lambda^2 xtc_2(0)\right|^{\frac{b}{\lambda}} \quad (66)$$

At a measurement event, $(x - c_0 t) = 0$, and $P(x = c_0 t)$ of finding a tachyon (as a photon) at the MP, is one, and so

$$P(x = c_0 t) = 1 = (c_1(0))^{\left(\frac{b}{\lambda}\right)}\left[1 - \lambda^2 c_2^2(0)t^2\right]^{\left(1+\frac{b}{\lambda}\right)} \quad (66A)$$

Since the left side is independent of $t$, $b = -\lambda$ at the MP. Eq.(66A) then says

$$c_1(0) = 1.$$

At points other than the MP, the probability density of a tachyon may be obtained from Eq.(66):

$$P(x,t) = \left[\{c_1(0)+\lambda x\}\{1-\lambda c_2(0)t\}\right]^{\left(1+\frac{b}{\lambda}\right)} \quad (67)$$

The total probability of finding a tachyon at all such points at time $t$ is

$$\int_{(\int cdt)}^{\infty} P(x,t)dx = [1-\lambda c_2(0)t]^{\left(1+\frac{b}{\lambda}\right)} \int_{(\int cdt)}^{\infty}[1+\lambda x]^{\left(1+\frac{b}{\lambda}\right)}dx$$

$$= [1-\lambda c_2(0)t]^{\left(1+\frac{b}{\lambda}\right)}\left[\frac{(1+\lambda x)^{\left(2+\frac{b}{\lambda}\right)}}{\lambda\left(2+\frac{b}{\lambda}\right)}\right]_{(\int cdt)}^{\infty}$$

$$= [1-\lambda c_2(0)t]^{\left(1+\frac{b}{\lambda}\right)}\left[\frac{1}{2\lambda+b}\left\{(1+\infty)^{\left(2+\frac{b}{\lambda}\right)} - \left(1+\lambda\int cdt\right)^{\left(2+\frac{b}{\lambda}\right)}\right\}\right].$$

To avoid the looming divergence in total probability it is imperative to have

$$2 + \frac{b}{\lambda} = 0, \quad \text{or,} \quad b = -2\lambda. \quad (67A)$$

The total probability for finding a tachyon anywhere at time t takes the form

$$\frac{0}{0}$$

which is undefined or meaningless. For tachyon, such a question of total probability in real spacetime has no meaning whatsoever ----- a result that is backed by the evidence that no tachyon has ever been detected. Eq.(67) now reads, with $\frac{b}{\lambda} + 2 = 0$ and $c_2(0) = c_1(0)c_2(0) = c_0$,

$$P(x,t) = \frac{1}{(1+\lambda x)(1-\lambda c_0 t)} \tag{68}$$

Since

$$c(x,t) = c_1(x)c_2(t) = \frac{c_2(0)(1+\lambda x)}{1-\lambda c_2(0)t} = \frac{c_0(1+\lambda x)}{1-\lambda c_0 t}$$

$$\frac{c_0}{c(x,t)} = \frac{1-\lambda c_0 t}{1+\lambda x} \tag{69}$$

The variation of fine structure constant is evaluated as

$$\frac{\Delta \alpha}{\alpha} = \frac{\alpha(x,t) - \alpha(0,0)}{\alpha(0,0)} = \left[\frac{c_0}{c(x,t)} - 1\right]$$

From Eq.(69) I find

$$\frac{\Delta \alpha}{\alpha} = -\left[\frac{\lambda(x+c_0 t)}{(1+\lambda x)}\right]. \tag{70}$$

In the above formulation of $\frac{\Delta \alpha}{\alpha}$, $\alpha$ is measured at two spacetime points (x , t) and (0,0). Generally, quantum measurement takes place at a single spatial point, say, $x_0$. Therefore, $\alpha$ is instead measured at spacetime points $(x_0, 0)$ and $(x_0, t)$. Hence

$$\frac{\Delta \alpha}{\alpha} = \frac{\alpha(x_0,t) - \alpha(x_0,0)}{\alpha(x_0,0)} = \left[\frac{c_1(x_0)c_2(0)}{c_1(x_0)c_2(t)} - 1\right] = [-\lambda c_0 t] = [-\lambda c_0 t_0 f] \tag{71}$$

if $t = ft_0$ is the look-back time, where f is the fractional look-back time and $t_0$ is the present time (=13.86 Gyr). What is the value of $\lambda$? It has a profound relationship with Quantum Gravity of photons.

# 6. QUANTUM GRAVITY OF PHOTONS, AND QUANTUM MECHANICS LEADS TO STRING THEORY

From Eq.(29a) one obtains

$$c_1^2(x)c_2^2(t)\left[\frac{1}{R_1}\frac{d^2R_1}{dx^2} - \kappa^2\right] = \frac{1}{R_2}\frac{d^2R_2}{dt^2} - \omega^2$$

or, $\quad \hbar^2\omega^2 = \hbar^2\kappa^2 c^2(x,t) + \frac{\hbar^2}{R_2}\frac{d^2R_2}{dt^2} - \frac{\hbar^2 c^2(x,t)}{R_1}\frac{d^2R_1}{dx^2}$

which yields the energy

$$E = \pm pc\left[1 + \left(\frac{\hbar^2}{p^2c^2R_2}\frac{d^2R_2}{dt^2} - \frac{\hbar^2}{p^2R_1}\frac{d^2R_1}{dx^2}\right)\right]^{\frac{1}{2}}.$$

Assuming that the quantity in parenthesis is small and considering only the positive energy

$$E = pc\left[1 + \frac{1}{2}\left(\frac{\hbar^2}{p^2c^2R_2}\frac{d^2R_2}{dt^2} - \frac{\hbar^2}{p^2R_1}\frac{d^2R_1}{dx^2}\right)\right]$$

one finally arrives at

$$E = pc + \frac{1}{2}\left[\frac{\hbar^2}{pcR_2}\frac{d^2R_2}{dt^2} - \frac{\hbar^2 c}{pR_1}\frac{d^2R_1}{dx^2}\right].$$

We restate it as

$$E = pc + V_Q \tag{72A}$$

where

$$V_Q = \frac{1}{2}\left[\frac{\hbar^2}{pcR_2}\frac{d^2 R_2}{dt^2} - \frac{\hbar^2 c}{pR_1}\frac{d^2 R_1}{dx^2}\right] \tag{72B}$$

is the quantum potential (of photon). Bohm's priceless contribution to Quantum Mechanics is perhaps the concept of quantum potential [38,39,40,41,42]. The quantum potential would generate a quantum force

$$F_Q = -\frac{\partial V_Q}{\partial x}$$

which has not been shown to be the fifth interaction in nature. But there is one possibility. Quantum Mechanics has successfully incorporated the electromagnetic, weak and strong forces into a successful model ----- the Standard model. But it could not do so with gravity. Since quantum force did not emerge as a separate interaction in the Standard model, could it be that gravity is already included in Quantum theory in the guise of quantum potential? There is a similar happening in string theory. In the quantum theory of strings, graviton appears in a natural way in the spectrum of closed strings. One more thing motivates : The constant $\lambda$ in the varying speed $c(x,t)$ is actually proportional to the Planck length, as will be shown below. This is an element of quantum gravity, that has played a central role in our theory. I, therefore label quantum potential $V_Q$ as quantum gravitational potential $V_G$ of photon.

From Eqs.(30a) and (30b),

$$\frac{1}{R_2}\frac{d^2 R_2}{dt^2} = (a+1)\omega^2$$

and,

$$\frac{1}{R_1}\frac{d^2 R_1}{dx^2} = (a+1)\kappa^2.$$

Inserting these in Eq.(72)

$$V_Q = \frac{1}{2}(a+1)\left[\frac{E^2}{pc} - pc\right] = \left(\frac{a+1}{2pc}\right)(E+pc)(E-pc) .$$

Since Eq.(72A) reads

$$E - pc = V_Q \tag{73}$$

$$V_Q = \left(\frac{a+1}{2}\right)\left(\frac{E+pc}{pc}\right)V_Q . \tag{73A}$$

Eq. (73) says that if $V_Q = 0$, we would be catapulted into classical regime. Hence, $V_Q \neq 0$. Eq.(73A) then yields

$$1 = \frac{1}{2}(a+1)\left(1 + \frac{E}{pc}\right)$$

which after a little rearrangement yields

$$\frac{E}{pc} = \left(\frac{1-a}{1+a}\right) = \left(1 - \frac{2a}{1+a}\right).$$

Therefore,

$$E - pc = -\left(\frac{2a}{1+a}\right)pc \tag{74}$$

Combining Eqs.(73) and (74),

$$V_Q = -\left(\frac{2a}{1+a}\right)\hbar\kappa(x)c(x,t). \tag{75}$$

Since $\kappa(x) = \frac{1}{c_1(x)}$ and

$$c_2(t) = \frac{c_0}{1 - \lambda c_0 t},$$

we finally obtain

$$V_Q = -\left(\frac{2a\hbar}{1+a}\right)\left(\frac{c_0}{1 - \lambda c_0 t}\right) \tag{76}$$

We shall shortly find that

$$1 - \lambda c_0 t \neq 0$$

so that $V_Q$ suffers no singularity!

In order to find the value of $a$ in Eq.(76), note that

$$\int \omega dt = \int \frac{c_0 dt}{1 - \lambda c_0 t} = -\frac{1}{\lambda}\ln|1 - \lambda c_0 t|, \tag{77}$$

also,

$$\int \kappa dx = \int \frac{dx}{c_1(x)} = \int \frac{dx}{1 + \lambda x} = \frac{1}{\lambda}\ln|1 + \lambda x| \tag{77}$$

whence follows the required constraints:

$$1 - \lambda c_0 t \neq 0$$

$$1 + \lambda x \neq 0.$$

From Eq.(30), with the input $b = -2\lambda$, the time-dependent probability density

$$R_2(t) = \frac{\sqrt{A}}{c_0}\sqrt{c_2(t)}.  \qquad (78)$$

From Eq.(78),

$$\frac{dR_2}{dt} = \frac{\sqrt{A}}{2c_0}\left(\lambda c_2^{\frac{3}{2}}\right)$$

whence

$$\frac{d^2 R_2}{dt^2} = \frac{3}{4}\lambda^2 \omega^2 R_2(t). \qquad (79)$$

Here we have made use of Eqs.(34a),(78) and $c_2(t) = \omega(t)$. To obtain the value of $a$ we consider Eqs.(79) and (29a) to find

$$a = \frac{3\lambda^2}{4} - 1 \qquad (80)$$

With this value of $a$ and Eq.(76), the quantum potential is

$$V_Q = 2\hbar\left(\frac{4}{3\lambda^2} - 1\right)c_2(t). \qquad (80a)$$

It is a reasonable assumption that $\lambda = \frac{dc_1}{dx}$ will be so small that $\frac{4}{3\lambda^2} \gg 1$.
Therefore, Eq. (80a) becomes

$$V_Q = \frac{8\hbar c_0}{3\lambda^2(1 - \lambda c_0 t)}. \qquad (81)$$

We further approximate by contending that $1 \gg \lambda c_0 t$ (the latter is of the order of $10^{-5}$) so that

$$V_Q = \frac{8\hbar c_0}{3\lambda^2}. \qquad (83)$$

I now derive the tachyon wave function from accumulated results. Setting $b = -2\lambda$, Eq.(67) becomes

$$P(x,t) = \frac{1}{[(1+\lambda x)(1-\lambda c_0 t)]}$$

and with help from Eq.(77) the tachyon wave function is

$$\psi(x,t) = \sqrt{P(x,t)} \exp\left[i\int \kappa dx - i\int \omega dt\right]$$

$$= (|1+\lambda x||1-\lambda c_0 t|)^{\left(-\frac{1}{2}+\frac{i}{\lambda}\right)} \tag{84}$$

I now digress a bit to General relativity and note that the Schwarzschild solution

$$c^2 d\tau^2 = \left(1 - \frac{2GM}{c^2 r}\right) c^2 dt^2 - \left(1 - \frac{2GM}{c^2 r}\right)^{-1} dr^2 - r^2(d\theta^2 + \sin^2\theta d\phi^2) \tag{85}$$

has a singularity at the Schwarzschild radius

$$R_s = \frac{2GM}{c^2} \,. \tag{85A}$$

Since the least possible invariant mass of photon is

$$M = m = \frac{\hbar}{2c_0} \tag{85B}$$

the Schwarzschild radius for a photon is

$$R_s = \frac{2GM}{c_0^2} = \frac{2G}{c_0^2}\left(\frac{\hbar}{2c_0}\right) = \frac{G\hbar}{c_0^3} = l_p^2 \tag{86}$$

where $l_p = \sqrt{\frac{G\hbar}{c_0^3}}$ is the Planck length! Eq.(86) demonstrates the union of

General relativity $\left(R_s = \frac{2GM}{c_0^2}\right)$ with Quantum Mechanics ( photon mass

$M = (\hbar/2c_0)$) leading to Quantum Gravity $\left(l_p = \sqrt{\frac{G\hbar}{c_0^3}}\right)$ of photons. The prediction

for Schwarzschild radius of photon is thus $2.61 \times 10^{-66} cm.$ Of special interest is the peculiar dimensional relation of Eq.(86) : $R_s = l_p^2$ means $cm. = (cm.)^2$, i.e.

$$cm. = 1$$

which implies that in Planck regime length melts into dimensionless numbers !

I now calculate the gravitational potential energy of a photon owing to back-reaction of its own gravitational field. If the photon is assumed to be a point particle, the calculated energy diverges owing to singularity at the origin. But this type of singularity that plagued QED is created by a belief of the existence of zero-distance. Thanks to the following theorem, zero-distance is quantum-mechanically forbidden by Heisenberg uncertainty relation.

*Theorem: Position of a quantum system cannot be measured with arbitrary precision.*

*Proof:* Let us consider a quantum system having a definite or precise position $x$. Then its uncertainty in position $\Delta x = 0$. Inserting this value in position-momentum uncertainty relation

$$\Delta x \Delta p_x \geq \frac{\hbar}{2},$$

one obtains $\hbar \leq 0,$
since division by zero is not a mathematically permissible operation. Therefore, $\Delta x \neq 0$. In other words, position of a quantum system cannot be measured with arbitrary precision. ◁
The same conclusion holds for its momentum. ◁

A quantum system cannot have a position at the origin ,i.e , $x = 0,$ because then its position uncertainty $\Delta x = 0$, which is not allowed according to the above theorem. Similarly, x = y = z = 0 ,and hence, zero volume are simply quantum-mechanically forbidden. Big bang singularity, owing to zero volume, is thus removed in Quantum Cosmology [37]. Since

$$\sqrt{x^2 + y^2 + z^2} = r = 0$$

implies x = y = z =0, the spacetime singularity at the center $r = 0$ of a Schwarzschild black hole is readily removed by Quantum Mechanics. The divergences occurring in Quantum Field Theory due to zero-distance interactions may be cured similarly. Like string theory (where strings collide over a non-zero distance), particles interact over a non-zero finite distance.

To determine the minimum distance between the field point (or MP) and the center of mass of a photon, note that the reference frame at the field point is generally at rest, i.e., stationary. But this scenario changes once the field point $x < R_s$ i.e., inside the photon's Schwarzschild radius $R_s$. No stationary frames are available inside that radius. Hence the minimum approachable distance between the field point in a *stationary frame* and the center of mass of a photon is $R_s$. The quantum gravitational potential energy of a photon placed in its own field is

$$V_G = -\frac{GM^2}{R_S} = -G\left(\frac{1}{l_p^2}\right)\left(\frac{\hbar^2}{4c_0^2}\right) = -G\left(\frac{c_0^3}{G\hbar}\right)\left(\frac{\hbar^2}{4c_0^2}\right) = -\frac{\hbar c_0}{4} \tag{86A}$$

where Eqs.(85B) and (86) have been utilized.

A few remarks about $V_Q$ of Eq.(83). Quantum potential has often been identified as an intrinsic self-energy of the quantum system [81]. R. Carroll [82] has shown that geometric properties of space, affected by the presence of a particle, is manifested as quantum force (via quantum potential). Photon mass $\left(\frac{\hbar}{2c_0}\right)$ introduces curvature (albeit small) of spacetime so it can attract objects gravitationally. This effect is obviously small. But it has recently been detected in a series of experiments showing a time-varying fine structure constant. I shall shortly prove that $\lambda$ is proportional to Planck length. Photon mass curves space, which in turn induces a space-time varying input in an otherwise constant speed of photon. The force at play is quantum gravity of photons ------ or, stated more transparently, ----- quantum self-gravity of photons. It is quite natural to identify the quantum potential $V_Q$ of a photon with the quantum gravity energy $V_G$, otherwise quantum force [83,84] stemming from quantum potential would admit the existence of an extra interaction other than the four fundamental interactions of nature. Conventional Quantum Mechanics with no explicit use of quantum potential, thus automatically excludes quantum gravity potential from the very beginning ! We need not add gravity to Quantum Mechanics. Gravity is included in Quantum Mechanics ! One cannot overlook the parallels between Quantum theory and string theory in this particular aspect. As Brian Greene[85] has said,`` String theory ……… is a quantum theory that includes gravity as well.`` Indeed Scherk and Schwarz [86] long ago found that string theory naturally includes gravitons as additional messenger-like particles in string vibrations.

Quantum potential has been shown to produce a local curvature of $\sqrt{P(x,t)}$ [87,88]. We thus have enough reason to equate $V_Q$ with $V_G$. Therefore, from Eqs.(83) and (86A)

$$\frac{8\hbar c_0}{3\lambda^2} = -\frac{\hbar c_0}{4},$$

whence

$$\lambda = \pm i 4\sqrt{\frac{2}{3}} \tag{87}$$

Eqs.(86B) and (87) express that the particle speed or momentum is imaginary. This is a pointer to the fact that the effect we are studying concerns tachyons. (I assumed tachyon mass real). The tachyon momentum

$$p = \frac{imv}{\sqrt{\frac{v^2}{c_0^2} - 1}}$$

becomes $p = imc_0$ when $\frac{v^2}{c_0^2} \gg 1$. The photon mass $\frac{\hbar}{2c_0}$ is carried over by tachyons in the quantum domain. This gives the tachyon momentum

$$p = i\left(\frac{\hbar}{2c_0}\right)c_0 = i\frac{\hbar}{2}.$$

We shall shortly show below that this tachyon momentum corresponds to an open string in its lowest excitation mode. Since the string vibrations are to and fro, the tachyon momentum is

$$p = \pm\hbar\kappa_0 = i\frac{\hbar}{2},$$

or, $\quad \kappa_0 = \mp\frac{1}{2}i$ (88)

It might seem implausible that

$$\kappa_0 = \left|\frac{\partial\phi}{\partial x}\right|$$

may be imaginary. But $\kappa_0$ is a vector which I define as

$$\vec{\kappa}_0 = \hat{x}'\kappa_0 = \hat{x}'\left(\mp\frac{1}{2}i\right) = \frac{1}{2}(\mp i\hat{x}').$$ (88A)

The string vibrations are along the imaginary axis $\mp i\hat{x}'$. Therefore, we have now entered two-dimensional space (one real and the other imaginary) and hence the minimum distance $l_p$ must be symmetric with respect to the real x-axis. The wavelength $\lambda_t$ corresponding to minimum tachyon momentum is thus $2l_p$. Then,

$$\kappa_0 = \frac{2\pi}{\lambda_t} = \frac{\pi}{l_p}$$

With Eq.(88) the above reads

$$(\pm i) = \frac{l_p}{2\pi} \tag{89}$$

Eqs.(87) and (89) yield the value of $\lambda$:

$$\lambda = 4\sqrt{\frac{2}{3}}(\pm i) = 4\sqrt{\frac{2}{3}}\left(\frac{l_p}{2\pi}\right) = \left(\frac{2\sqrt{2}}{\pi\sqrt{3}}\right)l_p \tag{90}$$

Eqs.(87) and (89) are puzzling in the sense that a real length is equated to an imaginary quantity. It was known long before that something peculiar happens at Planck scale. Since we are on the limit of least possible distance,to quote Edward Witten, ``space will re-expand in another `direction` peculiar to string theory.''[70]. Eq.(89) simply states that Planck unit of length is proportional to one unit of length along the imaginary spatial axis. It is known that inside the Schwarzschild radius (equal to Planck length squared), proper radial distances and proper times are imaginary.
   We have already found that the wave number corresponding to minimum momentum is (1/2) along the imaginary spatial axis. The Figs.(3) and (4) clearly show an open string in its least excitation mode.The vibration is confined within the distance $l_p$ ----- the minimum distance along the real line [90].Fig.(4) shows this open string whose open ends travel with velocity $c_0$. The replacement of point particles of Quantum Mechanics with strings resolves the incompatibility between Quantum Mechanics and General relativity.Quantum Mechanics thus leads to string theory via Eq.(88).
   A remark about Eq.(88A). In three-dimensional motion the vector $\vec{\kappa}_0$ may be represented as

$$\vec{\kappa}_0 = ai\hat{x}' + bj\hat{y}' + ck\hat{z}'$$

which is the vector part of quaternion[89].This offers a scope to extend the theory in terms of quaternionic Quantum Mechanics [35,36].
   I have said before that tachyons in complex spacetime manifold transform into photons at the measurement point(which is a real space point).This is equivalent to tachyon condensation [91,92,93,94,95,].Tachyons are permitted within the spectra of both open and closed bosonic strings.In tachyon condensation, a tachyonic field ----- usually a scalar field ---- acquires vacuum expectation value and reaches the minimum of potential energy. Ashoke sen [96] observed that open bosonic strings should end on a space-filling D25 brane, and the tachyon should be identified with the instability mode of the above brane. He

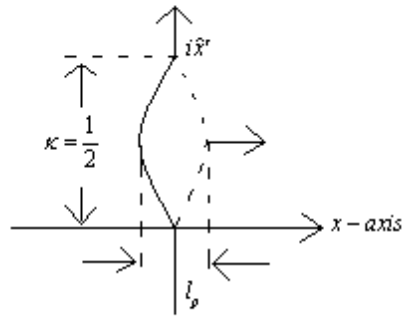
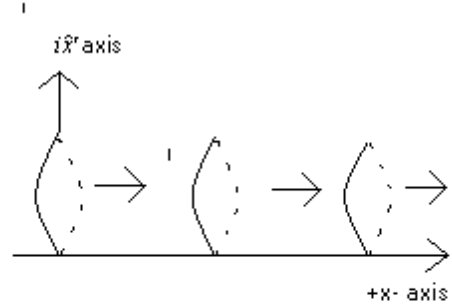

Fig.(3): An open string vibration corresponding to minimum momentum

Fig.(4): Open strings traveling with velocity $c_0$

conjectured that Witten`s open string field theory could be used to obtain a new vacuum in which the D25 brane is annihilated through condensation of tachyonic unstable mode. Open string tachyon condensation carries the physical system to a stable state, which, in our theory, is the photon at the measurement point where no tachyon exists.The Higgs mechanism [97] is an example of tachyon condensation. In our theory, the tachyon, in inaccessible complex spacetime, gives its mass to stable photon at real spacetime point. Sen`s three conjectures[57,58] triggered widespread investigation, and his first and third conjectures have been found to be almost accurately true.[59,60,61].Sen`s second conjecture may seem puzzling.Its first part says that ,after tachyon

condensation,the corresponding D-brane system disappears leaving nothing for open strings to end on. This is consistent with our finding that the condensation occurs at the MP, and the open string tachyon pops up as a photon at the MP. And in doing so, the open string curls up as a circle of radius $\pm i$, as is evident from Eq.(89): $l_p = \pm 2\pi(i)$. This implies that, at the MP, the tachyonic open string transforms into a closed circular string ----- which is consistent with the second part of Sen`s second conjecture.

Tachyon theories have always had the problem of causality violation staring in the face. But complex spacetime manifold comes to its rescue.In complex spacetime such phenomena would not occur in the first place. And tachyons ---- forever unobservable ----- introduces no pathological consequences to the physics at the measurement event.

It is sometimes said that tachyon speed increases as its energy decreases and vice-versa. This causes intriguing problems with charged tachyons. But I have assumed tachyon energy (and not tachyon mass ) as imaginary. Therfore, an accelerated charged tachyon loses imaginary part of tachyon speed. The real part of tachyon speed is zero. Tachyon speed is essentially imaginary on account of its imaginary momentum. This rules out a run-away effect of small acceleration generating large accelerations.

In Fig.(3) I have shown that the minimum approachable distance along the real axis is $l_p$ ---- the string is housed within its confines. Eq.(89) shows that

$$l_p = \pm 2\pi(i)$$

i.e. it is the circumference of a circle of radius $\pm i$. This confirms that the minimum approachable distance in Quantum Mechanics is not real. In other words,the real line ceases to exist at this point, i.e., there is a puncture of real space around the string. This result is quite similar to space-tearing flop transitions [71,72]. As the above equation shows, the hole at the puncture reveals the extra (imaginary ) dimension $l_p$ curled up into a circle of radius $\pm i$. It is not quite improbable to make this happen given the mind-boggling quantum gravity potential, on the order of $10^{53}$ Gev (obtained from Eqs.(83) and (90),in which the tachyon travels.

Non-zero photon mass may indicate Lorentz non-invariance[62,63,64,65,66,67,68].This is compatible with the fact that both string theory and loop quantum gravity allow for the possibility of violation of Lorentz symmetry. There are reasons to believe that Lorentz invariance may only be a low energy symmetry.This is suggested by divergences in Quantum Field Theory as well as Quantum Gravity theories. A useful way to study the Lorentz symmetry violation is through the possibility of a dispersion relation of the form [69] :

$$E^2 = p^2c^2 + m^2c^4 + Ap^2 + \frac{B}{\kappa_0}p^3 + ....$$

where A,B etc are constants. We find a similar dispersion relation containing only the $Ap^2$ term in Eq.(55a) :

$$E^2 = p^2c^2 + m^2c^4 + \left(\frac{c^4}{4\omega^2}\right)p^2 - \frac{\hbar^2 c^2}{4} \tag{90B}$$

An experiment devised to test Eq.(90B) may prove or disprove the violation of the classical concept of Lorentz invariance in Quantum theory.

Note that Lorentz symmetry is a postulate of the classical theory of special relativity. There is no scope of a preferred reference frame there. But the measurement problem of Quantum Mechanics has posed the problem of preferred basis[73]. In order to resolve it , one has to concede that the measurement frame (containing the MP) is hand-picked by the observer or measuring device [37]. The reference frame at the MP is thus a preferred frame.This idea is also supported by Copenhagen Interpretation.

Modern Quantum Field Theory does not exclude spontaneous breaking of Lorentz symmetry. A very small deviation from Lorentz invariance has been studied in the perspective of electro-weak interaction[74,75].

Robert Weingard had remarked,``It seems to me that one could be optimistic about a developing theory, even without such evidence, if there were at least some clues in either theory or experiment to suggest that the new theory was on the right track. Unfortunately, even this is lacking in string theory.''[70]. Now that Quantum Mechanics has been able to reveal the ground state (tachyon) of free bosonic string, it may now be arguably said that string theory is on the right track.

# 6. VERIFICATION OF THEORY WITH EXPERIMENTAL RESULTS

Various experiments have been carried out to find the tiny variation of fine structure constant [15,23,16,76,24,25,26,14,32,33,34]. I shall compare my results with those in the following references:

(A) M.T.Murphy et al (2002),astro-ph/0210532
(B) J.K.Webb et al (2002),astro-ph/0210531
(C) J.K.Webb et al (2001),astro-ph/0012539
(D) H.Marion et al, Phys.Rev.Lett.**90** ,150801(2003)
(E) S.Bize et al ,Phys.Rev.Lett.**90** ,150802(2003)

Inserting the value of $\lambda = \frac{2\sqrt{2}}{\pi\sqrt{3}} l_p = 0.52\ l_p$ in Eq.(71) I obtain

$$\frac{\Delta\alpha}{\alpha} = -\frac{2\sqrt{2}}{\pi\sqrt{3}} l_p c_0 t_0 f = -0.52 l_p c_0 t_0 f \tag{91}$$

where we set

$l_p$ = Planck length = $1.616 \times 10^{-33} cm.^{-1}$

$c_0 = 2.998 \times 10^{10} cm.\sec^{-1}$

$t_0 = 13.86 Gyr. = 4.37 \times 10^{17}$ sec.

f = fractional look-back time.

The unit of Planck length, $cm.^{-1}$ may seem bizarre, but it stems from the peculiar relation I have proved in Planck regime: cm.= 1, so $l_p$ cm.= $l_p$ cm$^{-1}$.

This makes $\frac{\Delta\alpha}{\alpha}$ in Eq.(91) dimensionless.

I first compare the theoretical results $\left(\frac{\Delta\alpha}{\alpha}\right)_{th} = V$ (for variation ) with the experimental findings of the paper (A) quoted above [lower panel of Fig.1(c)] In the following , EB stands for the solid error bar, and u = $\times 10^{-5}$.

[1] f = 0.4,   V = -0.439u, [falls in EB]
[2] f = 0.49,  V = -0.539u, [falls in EB]
[3] f = 0.6,   V = -0.66u,  [falls in EB]
[4] f = 0.66,  V = -0.726u, [matches almost exactly]
[5] f = 0.53,  V = -0.583u, [falls in EB]
[6] f = 0.56,  V = -0.616u, [falls in EB]
[7] f = 0.625, V = -0.688u [misses the EB]
[8] f = 0.71,  V = -0.781u [misses the EB]
[9] f = 0.745, V = -0.82u  [falls in EB]
[10]f = 0.77,  V = -0.847u, [matches exactly]
[11] f = 0.79, V = -0.869u, [falls in EB ]
[12] f = 0.809, V = -0.89u, [falls in EB]
[13] f = 0.84, V = -0.924u, [misses the EB]

10 predictions fall in Ebs.

Now I refer to the results in paper (B) quoted above[ lower panel of Fig (2)]:

[1] f = 0.43, V = -0.473u, [falls in EB]
[2] f = 0.50, V = -0.550u, [falls in EB]
[3] f = 0.54, V = -0.594u, [falls in EB]
[4] f = 0.57, V = -0.627u, [falls in EB]
[5] f = 0.60, V = -0.660u, [misses the EB]
[6] f = 0.63, V = -0.690u, [falls in EB]

[7] f = 0.66, V = -0.726u, [matches almost exactly]
[8] f = 0.70, V = -0.771u, [matches almost exactly the binned result]
[9] f = 0.75, V = -0.825u, [misses the EB]
[10] f = 0.77, V = -0.847u,[falls in EB]
[11] f = 0.79, V = -0.869u,[falls in EB]
[12] f = 0.81, V = -0.891u,[falls in EB]
[13] f = 0.84, V = -0.924u,[misses the EB]

In the above, 10 predictions fall within EBs.
Now I refer to the results in Fig.(1) of the paper (C ) quoted above.

[1] f = 0.45, V = -0.495u, [falls in EB]
[2] f = 0.54, V = -0.594u, [falls in EB]
[3] f =0.59, V = -0.649u, [falls in EB]
[4] f = 0.65, V = -0.716u, [misses the EB]
[5] f = 0.73, V = -0.803u, [almost matches]
[6] f = 0.77, V = -0.847u, [falls in EB]
[7] f = 0.82, V = -0.903u, [falls in EB]

In the above , 6 predictions fall within EBs.

Instead of astronomical evidences, one may look for a time-varying $\alpha$ in earth-bound experiments using atomic clocks. Transitions between two nearly degenerate states of atomic clock aspired an accuracy[32]:

$$\frac{1}{\alpha}\frac{d\alpha}{dt} = 10^{-18} / yr.$$

To find $\frac{d\alpha}{dt}$, note that

$$\frac{d\alpha}{dt} = \frac{e^2}{\hbar}\frac{d}{dt}\left(\frac{1}{c_1 c_2}\right) = \frac{e^2}{\hbar c_1}\left[-\frac{1}{c_2^2}\frac{dc_2}{dt}\right] = -\alpha\lambda c_2(t)$$

As done earlier, I approximate $c_2(t)$ to $c_2(0) = c_0$, take the unit of $l_p$ as $cm.^{-1}$ because we have proved the peculiar relation earlier that cm.= 1. Hence,

$$\frac{1}{\alpha}\frac{d\alpha}{dt} = -\lambda c_0 = -0.52 l_p c_0$$

which is in $\sec^{-1}$. After conversion to yr. the predicted result is

$$\frac{1}{\alpha}\frac{d\alpha}{dt} = -0.794 \times 10^{-15} / Yr.$$

This theoretical result agrees with the upper limit of $10^{-15}/Yr.$ obtained in recent atomic clock experiments in the papers (D) and (E) above[33,34].

# 7. CONCLUSION

When a time-varying constant was first announced it was almost unanimously admitted that this would have profound implications in almost all branches of physics. The present status of all this ``changing $\alpha$`` experiments is that these consistently point towards a revision in our thinking. A change in the value of $\alpha$ would violate the Equivalence Principle ------ the conceptual underpinning of General Relativity. It was also predicted that our concept of space and time would undergo radical change.

We have witnessed quite a few of such surprises. The first surprise is that one could hardly expect that a time-varying $\alpha$ is due to quantum gravity of photons. A precise quantum gravity formula for the time-varying fine structure constant has been derived :

$$\frac{1}{\alpha}\frac{d\alpha}{dt} = -\left(\frac{2\sqrt{2}}{\pi\sqrt{3}}l_p\right)\left(\frac{c_0}{1-\frac{2\sqrt{2}}{\pi\sqrt{3}}l_p c_0 t}\right)$$

After a little approximation, this result has been compared with the results of atomic clock experiments. The variation over cosmological time has also been derived :

$$\frac{\Delta\alpha}{\alpha} = -\left(\frac{2\sqrt{2}}{\pi\sqrt{3}}l_p\right)c_0 t,$$

it has been found to nicely agree with the results from a many-multiplet analysis of 128 quaser absorption systems.

The second surprise is quite dramatic : Quantum Mechanics contains gravity! Gravity is included in Quantum Mechanics in the guise of quantum potential. Quantum potential has long taken a back seat in discussions of quantum phenomena. But this innocuous potential is actually the quantum gravity potential of the quantum system concerned. The proof of this is : The formula for $\frac{\Delta\alpha}{\alpha}$ derived from quantum potential confirmed the quaser results.

The third and perhaps the greatest surprise is that Quantum Mechanics almost effortlessly leads to open string theory.The Schwarzschild radius of photon and Planck length satisfy the relation $R_S = l_p^2$.

The fourth surprise : In Planck regime length melts into mere numbers.Though this is quite hard to digest, dimensional analysis inexorably leads to this .But it was anticipated that space loses meaning at Planck scale.

To summarize the important results obtained,the wave function of photon was first derived by assuming a constant space-time derivative of phase, i.e., considering it to have constant speed and conserved energy.It is expected that the quantum potential for such a photon is zero,since we have assumed it massless.Hence quantum gravity is absent in that photon wave function.I have given enough reasons to refute the objections to a photon wave function. Contrary to what is widely believed I have explicitly found the form of photon`s position operator.

Next I have explored the possibility that the functions $\frac{d\phi_1}{dx}$ and $\frac{d\phi_2}{dt}$ need not necessarily be constants. This led to a photon wave function that yielded a space-time varying speed of light:

$$c(x,t) = \frac{c_0\left[1 + \frac{2\sqrt{2}}{\pi\sqrt{3}} l_p x\right]}{\left[1 - \frac{2\sqrt{2}}{\pi\sqrt{3}} l_p c_0 t\right]}$$

which satisfies a differential equation

$$\frac{\partial c}{\partial t} = c \frac{\partial c}{\partial x}.$$

I averted the criticism leveled at varying speed of light (VSL) theories by showing that quantum phenomena occur in complex dimensional spacetime manifold and light speed is actually a complex quantity. Therefore rulers and clocks of real value units cannot conjure up the the necessary changes in their units in complex spacetime.

I have shown that c (x, t ) is greater than today`s speed of light $c_0$. This brought tachyons in the physics of varying $\alpha$ ----- but with a difference. I have acknowledged a tachyon`s imaginary energy and momentum in special relativity ------- thereby ruling out their observation in the classical theory. But this procedure could afford a real-valued mass for tachyon. I have derived tachyon wave function and have shown that they travel in complex spacetime. This forever seals the fate of any experiment revealing a tachyon . It also help explain that causality is not violated. In the process, I have found the explicit form of the relativistic mass operator. Later it was found that tachyons transform into photons at the measurement event. This led to a search for the least possible mass of photon. This observation is also equivalent to tachyon condensation in open string theory where the tachyon reaches a minimum energy and a stable state (photon ).To find the photon mass I solved Klein-Gordon equation with an ansatz. This produced probability density waves traveling with the particle speed. The Klein-Gordon wave function refutes all objections that have been raised time and again against it. This solution predicts a photon mass of $1.7 \times 10^{-38} gm.$, which exactly equals the mass obtained in measurement of Gamma Ray Burst event. I have found the following quantum-mechanical mass-energy relation :

$$E = Mc^2 - \frac{\hbar^2}{8M}\left(1 - \frac{v^2}{c^2}\right)$$

in place of the classical equation

$$E = Mc^2 .$$

The above quantum analog of mass-energy relation may be tested experimentally to check the validity of the theory. Violation of Lorentz invariance owing to non-zero photon mass has been established by deriving a dispersion relation

$$E^2 = p^2 c^2 + m^2 c^4 + \left(\frac{c^4}{4\omega^2}\right) p^2 - \frac{\hbar^2 c^2}{4} .$$

     I have advanced some strong arguments in favor of identifying the self-gravitational potential of photon with quantum potential. And this worked, ------- the theoretical prediction of $\Delta \alpha / \alpha$ agreed quite nicely with various types of results obtained from quaser and atomic clock experiments. The Schwarzschild radius of photon, $R_S$ was found to relate to Planck length $l_p$ in the following way:

$$R_S = l_p^2 .$$

This realized the unification of gravity with Quantum Mechanics. A theorem has been proved stating that a quantum particle cannot have a definite position. Nor it can have a definite momentum. This powerful result from Heisenberg uncertainty relation immediately removes the singularities at the big bang in quantum cosmology, at the center of a Schwarzschild black hole and those of the Feynman diagrams of QED.
         It was not an involved process to show that Quantum Mechanics leads to String theory. The tachyons of minimum momentum are carried by open strings constituting a half wavelength in the direction of

imaginary axis. According to string theory, these strings end on D-branes. Sen has conjectured the instability of D-branes. I have found that his second conjecture is consistent with our finding that , at the measurement point an open string tachyon disappears leaving a stable system (photon ) in its place.It has been assured that negative norm states ------ dreaded in string theory are actually permitted by standard Quantum Mechanics[37].Quantum Mechanics also reveals a puncture in real space ----- equivalent to space-tearing flop transitions in string theory.The overall physics of Quantum Mechanics indicates that string theory is on the right track .

I sincerely thank K.Ghosh , S.G.Deb, B Ghosh, B Chakraborty, K Goswami, S Ghosh, S Chakraborty, P Basak, A Chatterjee, M Chatterjee , J Chatterjee, S Chatterjee, R Chatterjee , S Mukherjee and M Bhattacharya.